# Medium-Range Atomic Correlation in Simple Liquids. III. Density Wave Theory

Takeshi Egami[1,2,3] and Chae Woo Ryu[1,4]

[1]Shull-Wollan Center, Department of Materials Science and Engineering, University of Tennessee, Knoxville, Tennessee 37996, USA
[2] Department of Physics and Astronomy, University of Tennessee, Knoxville, Tennessee 37996, USA
[3]Materials Sciences and Technology Division, Oak Ridge National Laboratory, Oak Ridge, Tennessee 37831, USA
[4]Department of Materials Science and Engineering, Hongik University, Seoul 04066, Korea
Corresponding author: Takeshi Egami (egami@utk.edu)
ORCID identifier: T. Egami (0000-0002-1126-0276), C. W. Ryu (0000-0003-1588-4868)

**Abstract**

Elucidating the atomic structure of liquid and glass is one of the important open questions in condensed matter physics. In the conventional *bottom-up* approach one starts with focusing on an atom and the short-range order of its neighboring atoms, and the global structure is described in terms of overlapping local clusters of atoms as building units. However, this local approach fails to explain the strong drive to form the medium-range order which is distinct in nature from the short-range order. We propose an even-handed scheme with an additional *top-down* approach. In the top-down approach one starts with a high-density gas state and seeks to minimize the global potential energy through density waves. The local bottom-up and global top-down driving forces are not compatible, and the competition and compromise between them result in a final structure with the medium-range order. The density waves are pinned to atoms through the phase factors and amplitudes which reflect atomic dynamics. This even-handed approach provides a more intuitive explanation of the structure of simple liquid and glass and its relation to properties of liquid, such as viscosity and fragility.

## I. Introduction

Liquid is a condensed matter with density comparable to that of a solid. In liquid atoms are bound together by attractive interatomic potential, and their movements are highly correlated. Its structure is not totally random, characterized by local order. Determining and understanding such local order in the atomic arrangement in liquid and relating it to properties, including the glass transition, is an open and challenging question [1 – 5]. The atomic structure of liquid and glass is usually described by the pair-distribution function (PDF), $g(r)$, which depicts the distribution of distances between two atoms;

$$g(r) = \frac{1}{4\pi N \rho_0 r^2} \sum_{i,j} \left\langle \delta\left(r - |\mathbf{r}_i - \mathbf{r}_j|\right) \right\rangle, \qquad (1)$$

where $\mathbf{r}_i$ is the atomic position of the $i$-th atom, $N$ is the number of atoms in the system, $\rho_0$ is the average number density of atoms, and <….> denotes thermal and ensemble average [1, 3, 6]. The PDF describes only the two-body correlations, whereas properties of liquid and glass often depend on correlations of higher orders [7 – 9]. Nevertheless, the PDF is most widely used in description of the structure, because it can be determined with high accuracy by x-ray or neutron diffraction measurement through the Fourier-transformation of the structure function, $S(Q)$, where $Q$ is the momentum transfer of scattering [1, 3, 6].

The PDF shows many peaks indicating shell-like structures around each atom. The first peak describes the short-range order (SRO) in the nearest neighbor shell, whereas the peaks beyond the first peak depict the medium-range order (MRO). A common approach to explain the glass structure is to consider how the nearest neighbor atoms are configured, and then try to figure out how the local structures are stacked up to form the MRO beyond the nearest neighbors [1 – 3, 10 – 13]. However, this bottom-up, hierarchical approach does not always work, because the relationship between the SRO and the MRO is not so direct, particularly at low temperatures where atoms are strongly correlated. In Part I of the current series of papers we show that the SRO and the MRO exhibit distinct temperature dependences through the glass transition: The MRO freezes at the glass transition temperature, $T_g$, whereas the SRO does not [14]. In Part II [15] we explain the temperature dependence of the MRO coherence length in terms of the fluctuations away from the structurally coherent ideal glass state. In this work (Part III) we discuss the origin of the MRO by proposing an inverted top-down approach, by starting with a high-density gas state and considering the effect of the interatomic potential in reciprocal space, using the density wave approach [3, 16, 17] and the idea of the pseudopotential [18]. The bottom-up and top-down approaches are incompatible, and in our scheme the competition and compromise between the two driving forces produce the structure with the MRO.

The MRO is characterized by fairly regular oscillations of the PDF of which amplitude attenuates with distance as

$$|g(r)-1| \approx \frac{\exp(-r/\xi_s(T))}{r}, \qquad (2)$$

where $\xi_s(T)$ is the structural coherence length. The process of determining the PDF by experiment through the Fourier-transformation of $S(Q)$ is non-trivial, often compromised by termination errors originating from the limited range of $Q$ in diffraction measurement [6]. For this reason, the MRO has not received sufficient attention it deserves. However, advances in instrumentation, including the use of synchrotron radiation and electrostatic liquid levitation, significantly improved the accuracy of the PDF measurement, and the MRO following eq. (2) was found to extend much beyond several neighbor shells [19 – 23]. For instance, the PDF of $Pd_{42.5}Ni_{7.5}Cu_{30}P_{20}$ alloy liquid was found to have strong oscillations reaching beyond 20 Å, in spite of its complex chemistry [23]. This observation naturally raises a question on the origin of such a strong MRO. Moreover, $\xi_s(T)$ was found to obey the Curie-Weiss law for temperature dependence above $T_g$,

$$\frac{\xi_s(T)}{a} = C \frac{T_g}{T - T_{IG}} \quad (T > T_g), \qquad (3)$$

where $a$ is the average near neighbor distance, and $T_{IG}$ is the ideal glass temperature where $\xi_s(T)$ diverges in extrapolation and is negative for all metallic liquids we studied [23, 24]. Below $T_g$ the structure is frozen, and $\xi_s(T)$ remains constant. In Part II of the current series of papers [15] we explained this temperature dependence in terms of local density fluctuations away from the structurally coherent ideal glass state [23] defined by,

$$G_0(r) = G(r)\exp(r/\xi_s(T)), \tag{4}$$

where

$$G(r) = 4\pi r \rho_0 [g(r) - 1]. \tag{5}$$

It was found that $\xi_s(T)$ is closely related to dynamic properties of liquid such as viscosity [23] and fragility [25]. The fragility coefficient,

$$m = \left. \frac{d \log \eta(T)}{d(T_g/T)} \right|_{T=T_g}, \tag{6}$$

where $\eta(T)$ is viscosity, characterizes how quickly viscosity changes with temperature just above $T_g$ [26]. It was found that for a variety of liquids $m$ is proportional to the number of atoms in the coherence volume at $T_g$, $n_c = \rho_0 \left( \xi_s(T_g) \right)^3$ [25]. Because $n_c$ is a measure of the number of atoms which are correlated to each other, this relationship suggests a direct link between liquid fragility and liquid cooperativity. Now, the Curie-Weiss law, eq. (3), implies that there is a driving force to increase the structural coherence in liquid as temperature is lowered.

The purpose of the present work is to discuss the origin of this driving force to increase cooperativity in terms of formation of density waves. At present the increase in cooperativity is explained mainly in terms of mode-coupling [27-29], jamming [30, 31] or random-first-order transition [32-35]. The present approach explores another angle to see this phenomenon through energetics rather than through dynamics or geometry. For simplicity we primarily focus on simple liquids with spherical interatomic potentials, such as metallic liquids. In this paper we begin with summarizing the nature of the medium-range order as discussed in our previous publications, introduce the density wave theory and the concept of the pseudopotential, point out that the density wave state with the wavevector corresponding to the minimum of the pseudopotential is the ground state of the system, demonstrate that the simple density wave state reproduces a realistic density profile and the atomic dynamics, discuss the implications of the results in relation to dynamic cooperativity, dynamical heterogeneity, atomic pinning of the density wave, random-first-order-transition theory, hard-sphere model and jamming, and ends with concluding remarks. Some details are delegated to Appendices.

## II.  Nature of the medium-range order in liquid

Ornstein and Zernike (OZ) proposed a scheme to connect the SRO to the MRO through a self-consistency equation, and predicted eq. (2) [10]. Slightly different approaches were suggested later by others [3, 36, 37], but all these approaches are based on the idea that the MRO is a direct consequence of the SRO. However, as we discussed elsewhere [14, 15, 38] there are fundamental differences in nature between the SRO and the MRO. Even though the SRO and the MRO are related, they are sufficiently distinct so that the behavior of the MRO cannot be readily predicted from the SRO alone. For instance, $\xi_s(T)$ shows that the MRO freezes at $T_g$, whereas the first peak of the PDF, which describes the SRO, changes smoothly through the glass transition [14]. The mean-field approximation used in the OZ theory is justifiable at high temperatures. However, in the supercooled state atoms are dynamically correlated and cooperative [3], so that the accuracy of predicting the MRO by the OZ theory is questionable. For example, it predicts the Curie-Weiss behavior, eq. (3), in high-temperature approximation, but with a positive $T_{IG}$ [39].

The fundamental difference between the SRO and the MRO is that they correspond to different kinds of correlations. The SRO is represented by the first peak of the PDF which describes the distribution of distances from the central atom to the nearest neighbors. The number of atoms involved, the coordination number, $N_C$, is 12 – 14 for simple liquids with dense-random-packed (DRP) structure, such as metallic liquids [2, 3, 40], and is less for covalent liquids. However, at longer distances a PDF peak represents a much larger number of interatomic distances. The width of the higher-order peaks in the PDF is of the order of 1 Å, much wider than the typical phonon amplitudes which are of the order of 0.1 Å. Therefore, the peaks in the PDF beyond the first neighbors do not represent individual interatomic distances, but instead they describe more coarse-grained local density fluctuations [38]. In other words, the first peak describes the *point-to-point* correlations, but the MRO describes the *point-to-set* correlations [41]. For this reason, even though the PDF is a two-body correlation function, some features of the MRO reflect those of higher-order correlations; the derivative of the PDF depends on three-body correlation among two atoms separated by $r$ and the third atom at $r + dr$. As a consequence, the higher order peaks of the PDF decay with time more slowly than the first peak does. For instance, the distance-dependent decay time, $\tau(r)$, of the two-time correlation function, the Van Hove function (VHF), $G(r, t)$ [42], increases linearly with distance [43]. This is because $\tau(r)$ beyond the first neighbors describes the statistical timescale of local density fluctuations, rather than the timescale of atom-atom correlations. As we argued in Ref. 43, the number of atoms in the shell between $r$ and $r + dr$, $N_r dr$, increases with $r$ as $N_r = 4\pi r^2 \rho_0$, so the number fluctuation is proportional to $\sqrt{N_r}$, thus to $r$, resulting in the linear increase in $\tau(r)$ with $r$. Simulations at different dimensions proved this argument by showing,

$$\tau(r) = \tau_0 + \tau_r \left(\frac{r}{a}\right)^\chi, \quad \chi = \frac{d-1}{2}, \tag{7}$$

where $a$ is the nearest neighbor distance defined by the first maximum in the PDF and $d$ is dimensionality [43]. The same argument leads to the form of eq. (2) without the OZ theory [14]. Conversely, even though eq. (2) was first derived by OZ based upon their equation, the validity of eq. (2) does not justify the OZ theory.

For a simple liquid the oscillations in the PDF representing the MRO are approximately described by

$$g(r) - 1 \approx A_{MRO} \frac{a}{r} \sin(Q_{MRO} r + \delta_{MRO}) \exp\left(-\frac{r}{\xi_s}\right), \quad r > r_{cutoff}. \tag{8}$$

where $A_{MRO}$ is the amplitude of the MRO oscillation, $\delta_{MRO}$ is the phase factor which is small, and $r_{cutoff}$ is the position of the first minimum of the PDF beyond the first peak [14]. The MRO wavevector, $Q_{MRO}$, is close to the position of the first peak of $S(Q)$, $Q_1$, because the first peak of $S(Q)$ is largely determined by the MRO oscillation [25, 44]. These parameters describe the state of coarse-grained density fluctuations. Therefore, the MRO in the structure may be better described by local density $\rho(\mathbf{r})$, and its Fourier-transform, $\rho(\mathbf{q})$, which represents density waves [16, 17], rather than by the detailed individual atomic positions.

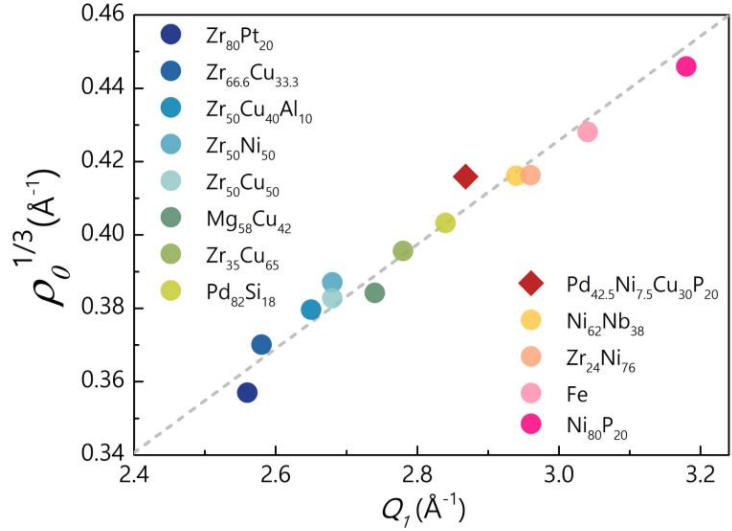

(a)

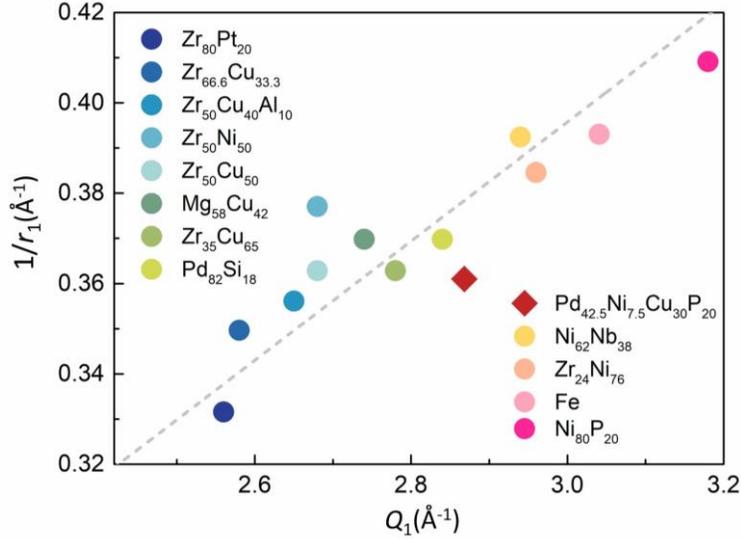

(b)

**Figure 1.** (a) Plot of $\rho_0^{1/3}$ against $Q_1$, the first peak position of $S(Q)$, for various metallic liquids by experiment for Pd$_{42.5}$Ni$_{7.5}$Cu$_{30}$P$_{20}$ alloy liquid and by simulation for others. The dashed line represents $Q_1 = C_{MRO}\rho_0^{1/3}$, with $C_{MRO}$ = 7.04 ± 0.023, and (b) plot of $1/r_1$ against $Q_1$, where $r_1$ is the first peak position of $g(r)$, for the same group of liquids. The dashed line represents $Q_1 = C_{SRO}/r_1$ with $C_{SRO}$ = 7.58 ± 0.062.

For various metallic liquids studied by simulation [23] $Q_1$ approximately shows a linear relation to $\rho_0^{1/3}$ as shown in Fig. 1 (a), by

$$Q_1 = C_{MRO}\rho_0^{1/3}, \text{ with } C_{MRO} = 7.04 \pm 0.023. \tag{9}$$

Therefore, the MRO periodicity normalized by $\rho_0^{1/3}$ is nearly universal to metallic liquids with the DRP structure, regardless of details of atomic bonds and interatomic potentials. In eq. (8) the first peak is at $a_{MRO} = 5\pi/2Q_{MRO} \approx 5\pi/2Q_1$ if we neglect $\delta_{MRO}$. If we assume this distance is twice the effective radius of an atom, $R_{eff}$, the volume fraction occupied by atoms by assuming a spherical atom, *i.e.* the atomic packing fraction (PF), is given by,

$$PF = \frac{4\pi}{3}\rho_0 R_{eff}^3 = \frac{125\pi^4}{48 C_{MRO}^3}, \tag{10}$$

which gives $PF = 0.726$ for the value of $C_{MRO}$ in eq. (9). This value is just below that of the *f.c.c.* structure (0.76) and of amorphous Fe (0.745 [45]), suggesting that the result in eq. (9) supports the idea that the DRP liquids have a nearly universal MRO periodicity which depends only on density. The position of the first peak of $g(r)$, $r_1$, is inversely related to $Q_1$ as shown in Fig. 1 (b) through $Q_1 = C_{SRO}/r_1$. The coefficient, $C_{SRO} = 7.58 \pm 0.062$, is consistent with the data on various metallic glasses [44]. The deviations in $C_{SRO}$, 0.82%, is almost three times larger than that of $C_{MRO}$, 0.33%. This is because $r_1$ is affected by local chemistry of SRO, whereas $Q_1$ reflects the average structure of the system.

The eq. (3) suggests that $\xi_s(T)$ diverges at $T_{IG}$, and the system reaches the state with long-range correlation without positional order, although such a state can never be achieved in reality, because $T_{IG}$ is negative and also liquid freezes at $T_g$. Nevertheless, it proves that there is a force to drive liquid to such a state at low temperatures. The $S(Q)$ of this state, the structurally coherent ideal glass state, has a $\delta$-function at $Q_{MRO}$ [23]. The three-dimensional $S(Q)$ is characterized by a Bragg sphere with the radius of $Q_{MRO}$. A quasicrystal [46] is the first example of a solid with long-range correlation without translational symmetry. It is a crystal in six-dimensions projected to three-dimensions with an irrational projection angle [47]. The structurally coherent ideal glass is a crystal in $N$-dimensions projected with nearly random phase factors, because the Bragg sphere can be considered to be made of infinite number of Bragg points covering the sphere. Interestingly, we were able to create a model with such features [23] using the Reverse Monte-Carlo (RMC) method [48] on $G_0(r)$ obtained by eq. (4) using the experimentally determined $G(r)$. In the present work we explain why this state obtained by extrapolation is energetically preferred. We use the density wave theory to discuss the origin of the driving force toward this state.

### III. Density wave picture of liquid structure

### A. Pseudopotential and density wave state

We will describe the structure in terms of local density, $\rho(r)$, and its Fourier-transform, $\rho(q)$, as

$$\rho(r) = \sum_i \delta(r - r_i) = \int \rho(q) e^{iq \cdot r} dq, \tag{11}$$

$$\rho(q) = \frac{1}{V}\int \rho(r) e^{-iq \cdot r} dr. \tag{12}$$

where $V$ is the system volume [16, 17]. The $\rho(q)$ describes the density waves and is related to the structure function, $S(q)$, by

$$S(q) = \frac{V}{\rho_0}|\rho(q)|^2. \tag{13}$$

For simplicity we consider a liquid in which atoms are interacting with a spherical interatomic potential, $\phi(r)$. The total potential energy is,

$$U = \frac{1}{V}\int \rho(r)\rho^*(r')\phi(|r-r'|)drdr' = \int |\rho(q)|^2 \phi(|q|)dq, \tag{14}$$

where

$$\phi(|q|) = \phi(q) = \int \phi(|r|)e^{-iq\cdot r}dr. \tag{15}$$

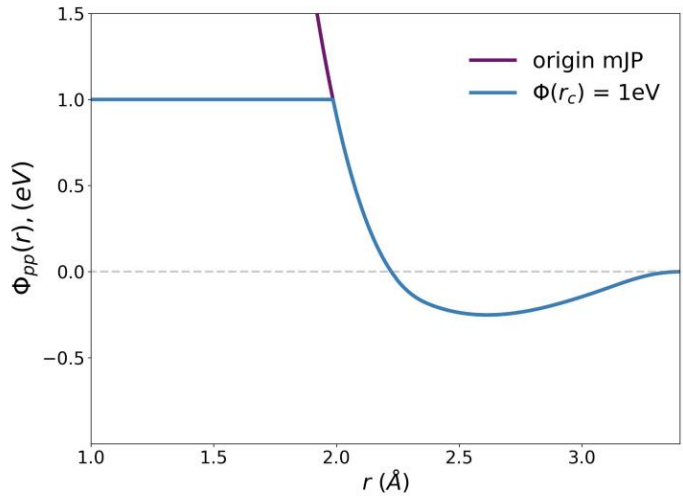

**Figure 2.** The modified Johnson potential (mJP) and $\phi_{pp}(r)$ of Fe with $\phi(r_c) = 1$ eV.

The $\phi(r)$ of Fe, the modified Johnson potential [49], is shown in Fig. 2. It has a minimum at $r_{min}$ = 2.62 Å, which is in close vicinity to the position of the first peak of the PDF, $r_1$. On the other hand, the $\phi(q)$ of Fe shows no minimum. However, this is misleading, because the $\phi(q)$ is dominated by the highly repulsive part of the $\phi(r)$, whereas the strongly repulsive part of the potential has no real effect on the total energy, because atoms cannot come so close to each other. For this reason, we split the potential into two parts,

$$\phi(r) = \phi_{pp}(r) + \phi_R(r) \tag{16}$$

Here $\phi_{pp}(r)$ is the "pseudopotential" in which the strongly repulsive part of $\phi(r)$ is removed and $\phi_{pp}(r) = \phi(r_c)$ for $r < r_c$ is assumed [18]. The cutoff, $r_c$, is chosen such that the cutoff temperature, $k_B T_u = \phi(r_c)$ is well above the actual temperature and no pair of atoms is found at distances at $r$

< $r_c$ (see Appendix A). The $\phi_R(r)$ is the repulsive part of the potential below $r_c$. Then the total potential energy is,

$$U = U_{pp} + U_R, \tag{17}$$

$$U_{pp} = \frac{1}{V}\int \rho(\mathbf{r})\rho^*(\mathbf{r}')\phi_{pp}(|\mathbf{r}-\mathbf{r}'|)d\mathbf{r}d\mathbf{r}' - U_{pp,self}, \tag{18}$$

$$U_R = \frac{1}{V}\int \rho(\mathbf{r})\rho^*(\mathbf{r}')\phi_R(|\mathbf{r}-\mathbf{r}'|)d\mathbf{r}d\mathbf{r}' - U_{R,self}. \tag{19}$$

Because $\phi(r)$ is defined between distinct atoms, we have to subtract the self-energies, $U_{pp,self}$ and $U_{R,self}$,

$$U_{pp,self} = \frac{1}{V}\int \rho(\mathbf{r})\rho^*(\mathbf{r}')W(|\mathbf{r}-\mathbf{r}'|)\phi_{pp}(|\mathbf{r}-\mathbf{r}'|)d\mathbf{r}d\mathbf{r}', \tag{20}$$

$$U_{R,self} = \frac{1}{V}\int \rho(\mathbf{r})\rho^*(\mathbf{r}')W(|\mathbf{r}-\mathbf{r}'|)\phi_R(|\mathbf{r}-\mathbf{r}'|)d\mathbf{r}d\mathbf{r}', \tag{21}$$

where,

$$W(z) = 1, \quad \text{if } z < R_{eff} = a/2 \\ \phantom{W(z) =} 0, \quad \text{if } z \geq R_{eff}, \tag{22}$$

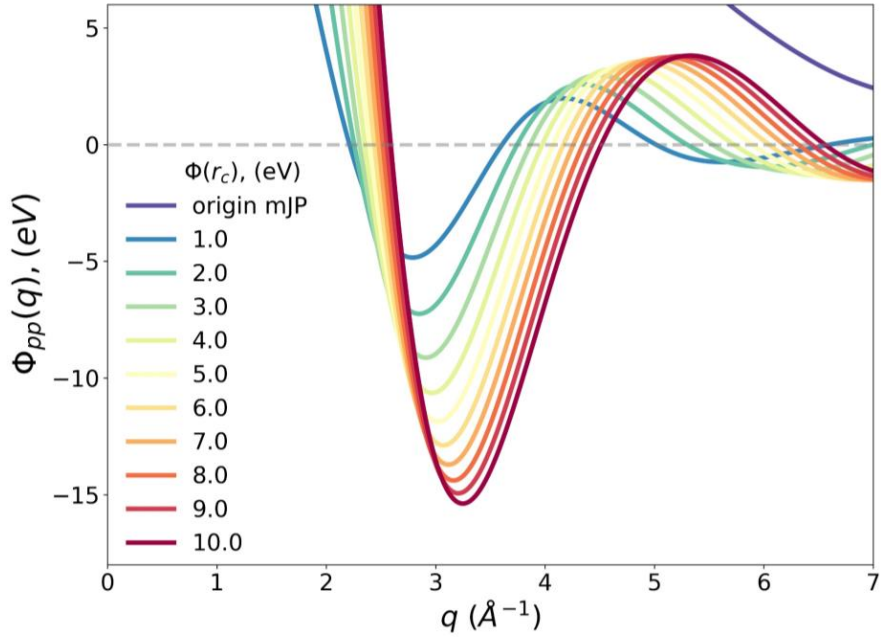

**Figure 3.** The $\phi(q)$ and $\phi_{pp}(q)$ of Fe with various cutoff values.

which defines the volume occupied by the central atom. Now, the ground state structure is determined by minimizing $U$. However, because no pair of atoms is found at distances at $r < r_c$ where $\phi_R(r)$ is non-zero, $U_R = 0$. Thus, it is sufficient to minimize $U_{pp}$ to determine the structure. This is equivalent to neglecting the high-energy, unreachable parts of the potential energy landscape (PEL) [4, 50] from consideration because they are irrelevant. For this purpose we divide $\rho(r)$ also into two parts,

$$\rho(\mathbf{r}) = \rho_{pp}(\mathbf{r}) + \rho_R(\mathbf{r}), \tag{23}$$

where $\rho_{pp}(r)$, the pseudo-density function, is chosen to minimize $U_{pp}$, whereas $\rho_R(r)$ is chosen to enforce $U_R = 0$.

Interestingly if the strongly repulsive part of the potential is removed the pseudopotential expressed in $q$-space, $\phi_{pp}(q)$,

$$\phi_{pp}(\mathbf{q}) = \phi_{pp}(|\mathbf{q}|) = \int \phi_{pp}(|\mathbf{r}|) e^{-i\mathbf{q}\cdot\mathbf{r}} d\mathbf{r} \tag{24}$$

shows a deep minimum at $q_{min}$, as shown in Fig. 3 for the modified Johnson potential with various cutoff energies, even though the original $\phi(q)$ has no minimum in the relevant range of $q$. If $\phi_{pp}(q)$ has a minimum at $\mathbf{q_{min}}$, $U_{pp}$, thus the potential energy $U$, is minimized by,

$$\rho_{min}(\mathbf{q}) = \rho_0 \delta(\mathbf{q} - \mathbf{q}_{min}), \quad \rho_0 = \frac{1}{V} \int \rho(\mathbf{r}) d\mathbf{r}. \tag{25}$$

This state is a long-range density-wave state with $\mathbf{q_{min}}$, akin to the structurally coherent ideal glass state (IG). Of course, to minimize the total energy the density waves with high-$q$s have to be involved as well, but the first minimum is so dominant that in the first order we just need to consider minimizing energy within the first valley of $\phi_{pp}(q)$. The value of $q_{min}$ depends only weakly on $r_c$, as shown in Fig. 4. At $\phi(r_c) = 4.5\ eV$ $q_{min}$ agrees with the wavevector for the structurally coherent ideal glass of Fe, $Q_{MRO} = 3.00\ \text{Å}^{-1}$, and with the position of the first peak of $S(Q)$, $Q_1 = 3.02\ \text{Å}^{-1}$. Therefore, it is reasonable to speculate that $\phi_{pp}(q)$ is driving the system to the structurally coherent ideal glass state. The $Q_1$ and $r_1$ are connected through the formula [44], $Q_1 r_1 \approx 5\pi/2$, reflecting the position of the first peak in eq. (8). Often the one-to-one correspondence between $Q$ and $r$ is assumed [51]. But this is highly misleading, because the first peak of $S(Q)$ corresponds to the MRO oscillations, whereas the first peak of the PDF affects the higher $Q$ part of $S(Q)$ [25, 44]. Therefore, it is most likely that the minimum in $\phi_{pp}(q)$ is providing the driving force to form the structurally coherent ideal glass state. There may be a way to define the optimum pseudopotential for which $q_{min} = Q_{MRO}$, but this is left for future study.

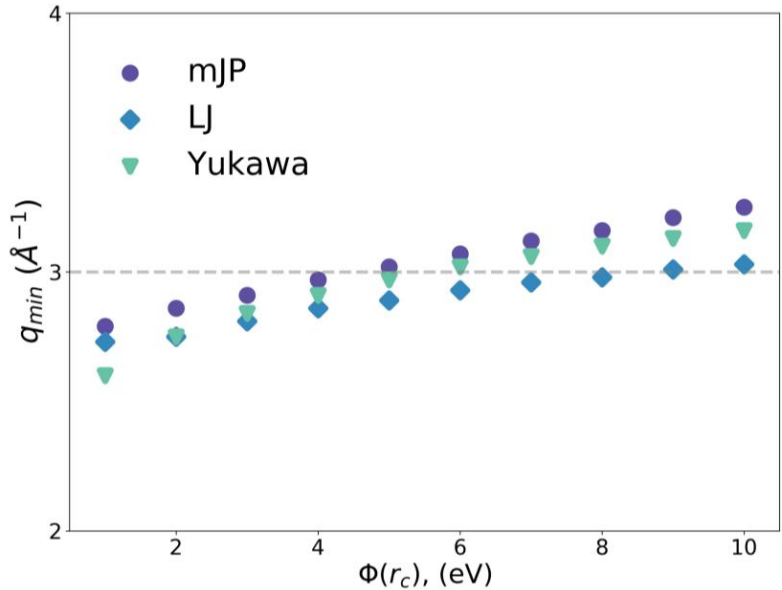

**Figure 4.** Dependence of $q_{min}$ on $\phi(r_c)$ for various potentials. The dashed line represents $Q_{MRO}$.

To examine the generality of the idea that the structurally coherent ideal glass state is driven by $\phi_{pp}(q)$, we studied $\phi_{pp}(q)$ for two other simple pairwise potentials, the Lennard-Jones (LJ) potential and the repulsive Yukawa (Y) potential. In both cases the parameters were chosen such that the atomic number density and $Q_1$ agree with those of liquid Fe (see Appendix B). As shown also in Fig. 4 the value of $q_{min}$ was found to be close to $Q_1$ and only weakly dependent on $\phi(r_c)$ for both cases. This applies even to the hard-sphere model [18]. Therefore, it is likely that in DRP liquids in general the structurally coherent ideal glass state is driven by $\phi_{pp}(q)$. Instead of focusing on $\phi(r)$ directly affecting two atoms at a time as in the bottom-up approach, we consider the effect of $\phi_{pp}(q)$ operating on the entire system of atoms in the high-temperature gaseous state in the top-down approach. Then, the minimization of the potential energy in eq. (18) by forming the density wave state is likely be the force to drive the liquid toward the structurally coherent ideal glass state. The presence of this substantial driving force toward the density waves explains why the well-defined oscillations characterizing the MRO persist up to long distances and high temperatures even for metallic alloy liquids with complex compositions.

## B. Structurally coherent ideal glass state

The structurally coherent ideal glass state is isotropic, and cannot be described by a single density wave, eq. (25). Instead, we consider the state defined by a large set of ideal glass $\boldsymbol{q_{IG}}$ vectors with the same length, $q_{IG} = |\boldsymbol{q_{IG}}| = Q_{MRO}$, evenly distributed over all angles in $\boldsymbol{q}$ space forming a Bragg sphere, and use it as the pseudo-density function,

$$\rho_{pp}(\boldsymbol{r}) = \rho_{IG,\Lambda}(\boldsymbol{r}) = \rho_0 + \int \rho_\Lambda(\boldsymbol{q_{IG}}) e^{i\boldsymbol{q_{IG}}\cdot\boldsymbol{r}} d\boldsymbol{q_{IG}}, \tag{26}$$

where,

$$\rho_\Lambda(\boldsymbol{q_{IG}}) = |\rho_\Lambda(\boldsymbol{q_{IG}})| e^{i\delta_{IG,\Lambda}(\boldsymbol{q_{IG}})}. \tag{27}$$

To keep the density real, we assume,

$$|\rho_\Lambda(-\boldsymbol{q_{IG}})| = |\rho_\Lambda(\boldsymbol{q_{IG}})|, \quad \delta_{IG,\Lambda}(\boldsymbol{q_{IG}}) = -\delta_{IG,\Lambda}(-\boldsymbol{q_{IG}}). \tag{28}$$

The phase factor, $\delta_{IG,\Lambda}(\boldsymbol{q_{IG}})$, is added to avoid atom pileups. It varies strongly and nearly randomly for the $\boldsymbol{q_{DW}}$ vectors in different directions. The sets of the phase factors, $\{\delta_{IG,\Lambda}(\boldsymbol{q_{IG}})\}$, and amplitudes, $\{|\rho_\Lambda(\boldsymbol{q_{IG}})|\}$, characterize each density wave state. For simplicity we use $\Lambda$ to designate the sets of the phase factors, $\{\delta_{IG,\Lambda}(\boldsymbol{q_{IG}})\}$, and amplitudes, $\{|\rho_{\Lambda}(\boldsymbol{q_{IG}})|\}$. Because there are infinite sets of these factors, there are infinite density wave states, $\rho_{IG,\Lambda}(\boldsymbol{r})$. Below, we show that this state, described by the pseudo-density function $\rho_{pp}(\boldsymbol{r})$, is close to the structurally coherent ideal glass state of liquid except for details of the SRO.

The $S(q)$ of the structure defined by eq. (26) has just the Bragg peak at $q_{IG}$, which is the first peak of the total $S(q)$. Because the first peak of $S(q)$ describes the MRO [25, 44], the PDF of this state represents just the MRO portion of the PDF,

$$g_{MRO}(r) - 1 = B_\rho(q_{IG}) \frac{q_{IG}^2}{\pi \rho_0^2} \frac{\sin q_{IG} r}{q_{IG} r}. \tag{29}$$

where

$$B_\rho(q_{IG}) = L^2 \left\langle |\rho(\boldsymbol{q_{IG}})|^2 \right\rangle, \tag{30}$$

and $V = L^3$. Its derivation is given in Appendix C. The total potential energy, eq. (17), is independent of the phase factors. Thus, all the density wave states characterized by the sets of the phase factors are degenerate in energy for a given value of $B_\rho(q_{IG})$. This basic degeneracy is reflected in high degeneracy of the basins in the PEL [4, 50].

The wavelength of this density wave in liquid Fe, $\lambda_{IG} = 2\pi/q_{IG} = 2.09$ Å, is too long to specify the atomic position with sufficient accuracy. In addition, because $\phi(r)$ is repulsive at short distances, $\rho(r)$ has gaps between atoms where $\rho(r) = 0$. The pseudo-density function, $\rho_{pp}(\boldsymbol{r})$, cannot accurately describe these details, even though it catches the overall structure. To describe the atomic positions and these gaps accurately we need to include $\rho_R(r)$, which represents contributions from waves with $q$s higher than $q_{IG}$ and eliminates atomic overlaps. Indeed, the $S(q)$ of the structurally coherent ideal glass state produced by the RMC shows features at higher $q$ [23]. In this sense $\rho_R(r)$ represents the SRO driven by $\phi_R(r)$, whereas $\rho_{pp}(r)$ describes the MRO driven by $\phi_{pp}(q)$. As we discuss below the competition between these two determines the final structure.

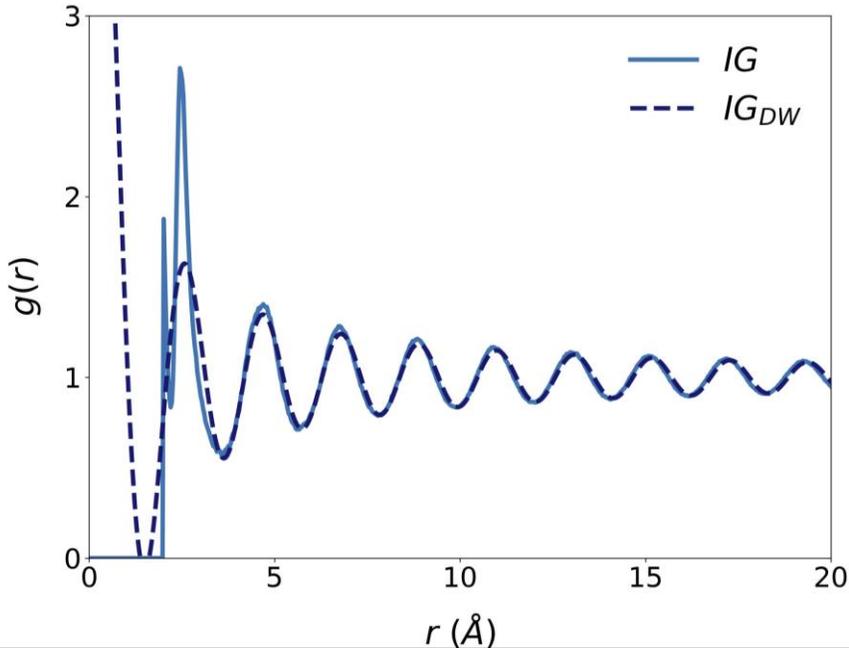

**Figure 5.** The PDF of the model of ideally coherent glass state for liquid Fe created by the RMC method (solid curve). The sharp peak at 2.0 Å reflects the imposed minimum atomic distance. The dashed curve represents eq. (29).

To test this density wave picture, we start with the model of ideally coherent glass state created by the RMC method based on the structure of liquid Fe of which PDF is shown in Fig. 5. The original model consists of 54000 atoms in a supercell of $L_s^3$ ($L_s = 88.3$ Å) with cubic periodic boundary conditions. However, because the cubic boundary conditions are inconsistent with the assumed isotropic nature of the density wave model, the model could not maintain correlation over the whole volume and the oscillations in $G(r)$ decayed beyond 50 Å. Thus, we took a central cubic portion of the model with $L = 50$ Å for the analysis of the structure and the PDF. The allowed $q$ vectors are, $q_n = \Delta q n$, $n = (n_x, n_y, n_z)$, where $\Delta q = 2\pi/L$ and $n_x, n_y, n_z$ are integers. The $q(n_{DW}) = \Delta q n_{DW}$ is chosen so that

$$\left(n(q_{IG})-\frac{1}{2}\right)^2 < n_{IG}^2 < \left(n(q_{IG})+\frac{1}{2}\right)^2, \quad n(q_{IG})=\frac{q_{IG}}{\Delta q}. \tag{31}$$

Then, the density is given by,

$$\rho_{IG}(r) = \rho_0 + \sum_{n_{IG}} \left|\rho(q(n_{IG}))\right| e^{i(q(n_{IG}) \cdot r + \delta_{IG}(q(n_{IG})))}. \tag{32}$$

The density profile of the $\rho_{IG}(r)$ by eq. (32) compares well except for details with that of the model, $\rho_{model}(r)$, as shown in Fig. 6. The $\rho_{model}(r)$, initially composed of $\delta$-functions, is convoluted with a Gaussian profile with the width $w$ (= 0.4 Å) to model the uncertainty in position, which is of the order of $\lambda_{IG}/2$, and mitigate the need of higher $q$ contributions. The PDF predicted by eq. (29) agrees well with the MRO portion of the PDF as shown in Fig. 5 by a dashed curve. As indicated by eqs. (29) and (30) the values of $|\rho(\boldsymbol{q}(\boldsymbol{n}_{IG}))|$ scale as $L^{-1}$. The $L|\rho(\boldsymbol{q}(\boldsymbol{n}_{IG}))|$ shown in Fig. 7 (a) is widely distributed in magnitude with the approximate probability of the two-dimensional Gaussian distribution reflecting the two-dimensional nature of the Bragg sphere,

$$P\left(L|\rho(\boldsymbol{q}(\boldsymbol{n}_{IG}))|\right) = \frac{L|\rho(\boldsymbol{q}(\boldsymbol{n}_{IG}))|}{2\pi\sigma^2_{\rho(q)}} \exp\left(-\frac{L^2|\rho(\boldsymbol{q}(\boldsymbol{n}_{IG}))|^2}{2\sigma^2_{\rho(q)}}\right). \tag{33}$$

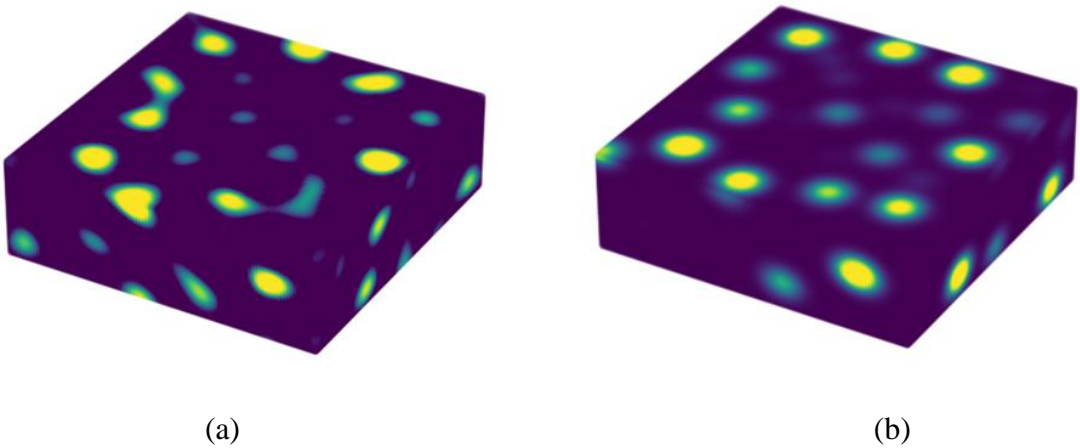

(a)          (b)

**Figure 6.** (a) Density profile of the $\rho_{IG}(r)$ by eq. (32) compares well with (b) that of the model, $\rho_{model}(r)$.

The phase factor, $\delta_{IG}(\boldsymbol{q}(\boldsymbol{n}_{IG}))$, is evenly distributed from $-\pi$ to $\pi$ as shown in Fig. 7 (b).

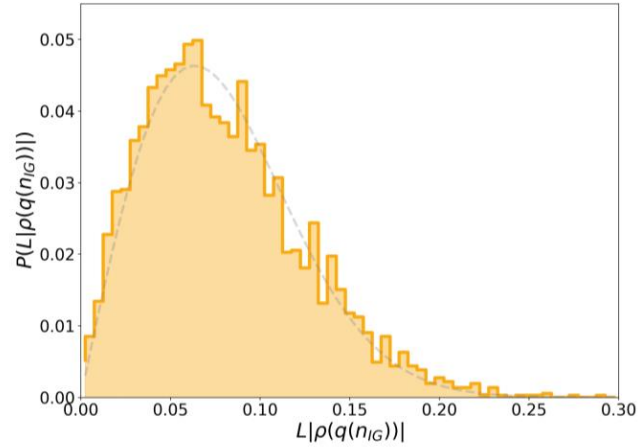

(a)

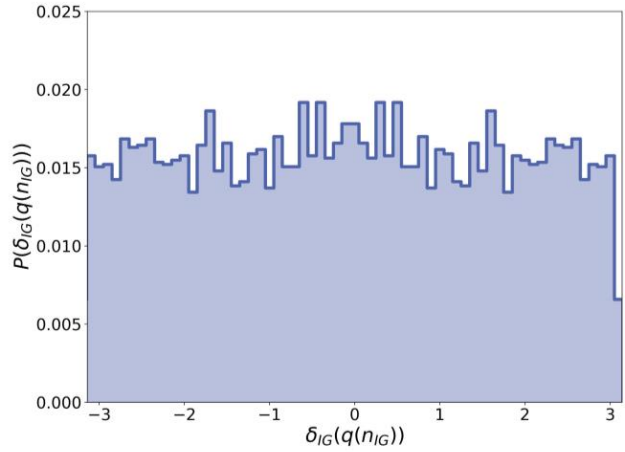

(b)

**Figure 7.** Distribution of values of (a) $L|\rho(\boldsymbol{q}(\boldsymbol{n}_{IG}))|$ and (b) $\delta_{IG}(\boldsymbol{q}(\boldsymbol{n}_{IG}))$. The dashed curve in (a) represents eq. (33).

## IV. Density waves in real liquid and glass

### A. MRO and density wave state

Whereas the $S(Q)$ of the structurally coherent ideal glass state has a Bragg-like first peak, those of real liquid and glass have a broader peak indicating limited spatial coherence in the structure characterized by the MRO. We now discuss why the structurally coherent ideal glass state cannot be achieved in reality. Note that the first peak of $g_{MRO}(r)$ in the density wave state calculated by eq. (29) is as wide as the second peak (Fig. 5). However, such a broad peak is not compatible with $\phi(r)$ which prefers the first peak as narrow as possible. Also, the wide first peak of $g_{MRO}(r)$ results in a very diverse local structures with the probability of the icosahedral clusters below 1% [23] in the structurally coherent ideal glass state, whereas icosahedral local structure is strongly preferred in the bottom-up approach [11-13]. Therefore, the driving force to produce the structurally coherent ideal glass state with $q = |\boldsymbol{q}_{IG}|$ is in conflict with the force to produce the SRO with as small local distortion as possible. In order to produce a model with a sharp PDF first peak showing better SRO we have to involve wider ranges of density waves, whereas the density wave model, eq. (26), uses only one $q = |\boldsymbol{q}_{IG}|$. The conflict between the two driving forces, that the force toward a narrow $g(r)$ peak in real space by $\phi_R(r)$ and that toward a narrow $\rho(q)$ peak in reciprocal space by $\phi_{pp}(q)$, is similar to the conflict behind the quantum mechanical uncertainty principle between position and momentum, and it is a fundamental property of the Fourier-transformation. The final structure with the MRO emerges from the compromise of the competition between these two driving forces. This compromise is the mechanism behind the misfit strain, $\varepsilon_V^{T,mf}$, in Part II of this series [15]. At non-zero temperatures not only the density waves at $\boldsymbol{q}_{IG}$, but those over a range of $\boldsymbol{q}$ near $\boldsymbol{q}_{IG}$ are excited, resulting in a structure with even more limited MRO coherence with a wider first peak of $\rho(q)$. In the theory of the structural coherence in the MRO [15, 24], the long-range structural coherence in the ideal state is lost in real liquid because of thermally excited local distortions in the atomic cage and intrinsic structural frustration which prevents the system from reaching the structurally coherent ideal state.

Thus, in real liquid and glass the structural coherence is very much limited in space. Such a state can be modelled by a collection of a very large number of different ideal glass states, each characterized by $\Lambda$, confined to a limited volume. This spatial confinement results in broadening of the first peak of $S(Q)$. Then, the density profile of real liquid and glass is given as an assembly of the ideal states by,

$$\rho(\boldsymbol{r}) = \frac{1}{V} \sum_{\Lambda} w_{\xi,\Lambda}(\boldsymbol{r},\boldsymbol{r}') \rho_{IG,\Lambda}(\boldsymbol{r}), \tag{34}$$

where $w_{\xi,\Lambda}(\boldsymbol{r}, \boldsymbol{r}')$ is a window function with the size factor of $\xi$, centered at $\boldsymbol{r}'$, for each ideal glass state $\Lambda$. The variation with position in the base density wave state, $\Lambda$, represents the fundamental incoherence in the state of real liquid and glass. The window functions can be overlapping in space. In our model they are strongly overlapping exponentially decaying functions centered on each atom. If $w_{\xi,\Lambda}(\boldsymbol{r}, \boldsymbol{r}')$ is the same for all $\Lambda$ and dependent only on $\boldsymbol{r} - \boldsymbol{r}'$, through the convolution theorem the Fourier-transform of eq. (34) is,

$$\rho(\boldsymbol{q}) = \sum_{\Lambda} \rho(\boldsymbol{q}_{IG,\Lambda}) W_{\xi,\Lambda}(\boldsymbol{q} - \boldsymbol{q}_{IG}), \tag{35}$$

where $W_{\xi, \Lambda}(q)$ is the Fourier-transform of $w_{\xi, \Lambda}(r)$. Because $\rho(q_{IG,\Lambda})$ is a δ-function the peak profile of $\rho(q)$ is determined by $W_{\xi, \Lambda}(q)$. Therefore, by examining the shape of the first peak of $S(Q)$ we gain knowledge on the nature of the window function. For many good metallic liquids and glasses with long $\xi_s$, it is close to the Lorentzian, consistent with the window function with the exponentially decaying function, whereas it is closer to the Gaussian for liquids and glasses with short $\xi_s$. The term "ideality" was introduced to describe the closeness to the Lorentzian shape [52].

### B. Dynamics of density wave

In the density wave picture, there are two ways to describe the movement of atoms. The first is to change the amplitude of the density waves, $|\rho(q_{DW})|$ (amplitudon), and the second is to change the phase factor, $\delta_{DW}(q_{DW})$ (phason). The concept of phason is widely used in density wave theory [53]. In liquid they reflect atomic dynamics and fluctuate with time. To characterize such dynamic fluctuations, we calculated the autocorrelation functions for amplitude and phase,

$$C_{amp}(t) = \langle |\rho(q_{IG},0)||\rho(q_{IG},t)|\rangle - \langle |\rho(q_{IG},0)|\rangle^2, \tag{36}$$

$$C_{ph}(t) = \langle \delta(q_{IG},0)\delta(q_{IG},t)\rangle, \tag{37}$$

for the model of liquid Fe ($T_g \sim 950$K) at various temperatures. Note that $\langle \delta(q_{DW})\rangle = 0$. Both of them were found to decay exponentially with single relaxation times, $\tau_\rho$ and $\tau_\delta$. The temperature dependences of $\tau_\rho$ and $\tau_\delta$ are shown in Fig. 8, down to about 1.5 $T_g$. Crystallization prevented from studying the dynamics at lower temperatures. They are nearly identical and also agree except for constant factors with $\tau_{LC}$, the time for an atom to lose one neighbor [54], and with the decay time of the first peak of the Van Hove function [42], $\tau_{VH}$, which is proportional to $\tau_{LC}$ [55]. For metallic liquids above $T_A$ (~ $2T_g$), it has been shown that the Maxwell relaxation time ($\tau_M = \eta/G_\infty$, where $\eta$ is viscosity and $G_\infty$ is the high-frequency shear modulus) is equal to $\tau_{LC}$ [55]. The Arrhenius temperature, $T_A$, is a temperature above which the viscosity of a liquid obeys Arrhenius law [26]. Therefore, the dynamics of phason and amplitudon is directly related to the atomic dynamics at high temperatures. That $\tau_\rho \approx \tau_\delta \approx 2\tau_{LC}$ suggests that amplitudon or phason excitation incurs two bond cutting or forming actions. Similarly, $\tau_{VH} \approx 3\tau_{LC}$ implies that three bonds need to be cut for the peak of the Van Hove function to decay to $1/e$.

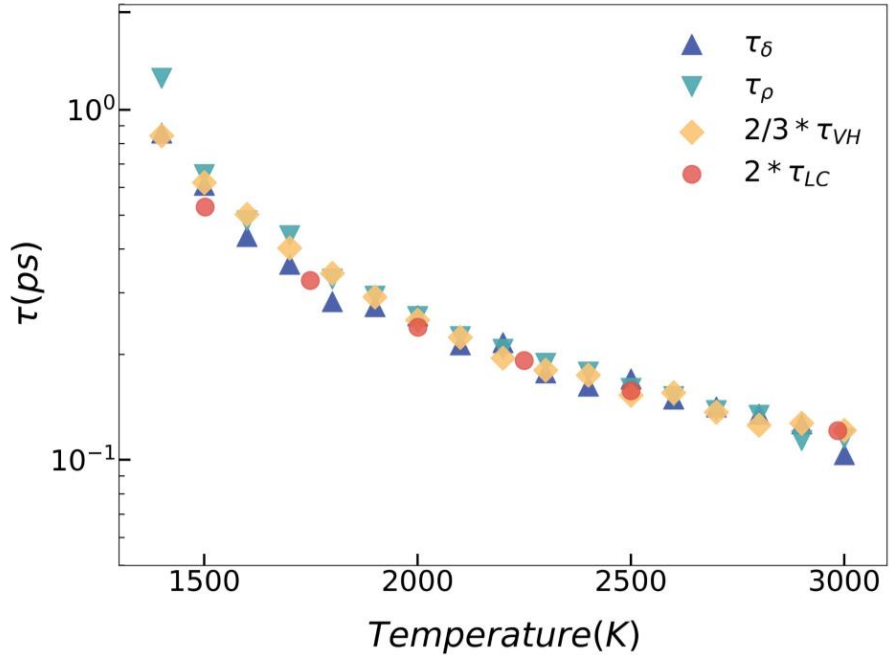

**Figure 8.** Temperature variation of the lifetime of $C_{amp}(t)$, $\tau_\rho$, and $C_{ph}(t)$, $\tau_\delta$, compared to $2\tau_{LC}$ and $(2/3)\tau_{VH}$ for liquid Fe.

## V. Discussion

### A. Dynamic cooperativity and the MRO

Below $T_A$ the viscosity of a liquid becomes super-Arrhenius and increases rapidly with decreasing temperature [26]. Various theories were proposed to explain this rapid rise in viscosity [23 – 31], but the issue is still open. It is widely agreed upon that the rise in viscosity is related to increase in atomic cooperativity [56 – 58]. The extent of cooperativity can be directly observed by the ratio, $\tau_M/\tau_{LC}$. Above $T_A$ $\tau_M/\tau_{LC} = 1$, implying that cooperativity is absent, and viscosity is determined by a single action of bond-cutting or bond-formation. This is because the timescale for structural fluctuation is shorter than the time for a phonon to travel from one atom to the next, so that atoms cannot communicate through phonons [54, 59]. However, below $T_A$ local actions of bond-cutting or bond-formation affect atoms nearby, causing cooperative dynamics, seen by the ratio $\tau_M/\tau_{LC}$ increasing above unity with decreasing temperature [54], indicating many bonds have to be severed for viscous flow to occur.

Bouchaud and Biroli [8] and Montanari and Semerjian [60] argued that the activation energy for viscosity in cooperative liquid and glass is proportional to the volume defined by the cooperative regions. Now, the region of atomic cooperativity has long been considered to be a local defect-like region where atomic density is low and atoms are easier to move [56 – 58, 61 – 65]. However, this assumption may need to be scrutinized in light of recent results. In crystalline solids structural defects provides atomic mobility, and this knowledge prompted the definition of defects in glasses as soft or loose regions. But there is a critical difference between defects in crystal and those in glass. In crystalline solids defects are defined in terms of local deviations from the lattice periodicity. When they move their identity does not change because they are topologically protected. On the contrary, in liquid and glass defects are not topologically protected, and change their structure and nature significantly upon motion. In fact, application of small strain significantly changes the activation energy spectrum for local deformation [66]. Furthermore, at the saddle point of the PEL the local structure configurationally melts for a short time (~ 1 ps) [67] and largely loses the memory of their thermal history [68].

In general, the saddle point is known as a generator of chaos [69, 70], and it is difficult to predict the behavior after passing through the saddle point. Even before the final saddle point is reached the local PEL surface is far from smooth and shows complex fractal heterogeneity [71, 72], because atomic connectivity changes all the time even in the apparently elastic regime [73, 74]. These numerous small local saddle points in the PEL can generate chaotic uncertainty, eventually wiping out previous history. Then, the saddle point height, which represents the activation energy and the resistance to motion, may not depend so much on the starting point of deformation. Even though deformation event may indeed be initiated by defects, the local structure changes significantly during the deformation event, so that the saddle point height may be determined by global statistics, such as the effective temperature [58, 75], rather than the initial state of flow. This argument explains why viscosity and fragility depends on the MRO coherence length, $\xi_s(T)$ [23, 25], which is a bulk property and is unrelated to defects.

### B. Dynamical heterogeneity

Supercooled liquid shows strong dynamical heterogeneity [7, 8, 61, 76 – 78]. The origin of dynamical heterogeneity is often ascribed to structural heterogeneity, particularly that of local softness [7, 8, 61 – 65, 76 – 78]. However, the structure with the MRO discussed here does not show strong structural heterogeneity. In fact, the structure is statistically homogeneous, with similar strongly overlapping exponential window functions in eq. (34). It is possible that dynamical heterogeneity, such as the string-like motion [76, 77], does not originate from structural heterogeneity, but is caused by cascade events of deformation [79]. In other words, dynamical heterogeneity may be a consequence of non-linear response of the media to stress, rather than that of structural heterogeneity. This point warrant further study.

Dynamical heterogeneity is characterized by the four-point correlation functions [7]. It is widely assumed that the two-point correlation functions, such as the Van Hove function, cannot describe dynamical heterogeneity. However, a recent simulation suggests that in certain cases the four-point correlation functions carry practically the same information as the Van Hove function does [80]. In the definition of the four-point correlation function, $\chi(r, t)$, a window function is introduced to eliminate the effect of thermal fluctuation [7]. Now, the MRO portion of the Van Hove function describes a point-to-set correlation, in which the effect of thermal fluctuation is already suppressed. As we discussed above, the decay time of the Van Hove function, $\tau(r)$, increases linearly with $r$, resulting in de Gennes narrowing [43], because they describe the decay of the density correlations, rather than that of the two-point atomic correlations. Therefore, it is possible that the MRO portion of the Van Hove function, which has not received sufficient attention it deserves, carry some information on dynamical heterogeneity.

A recent simulation work [81] show that the dynamics of the four-point correlation function is directly related to the dynamics of speckle patters observed by the x-ray photon correlation spectroscopy (XPCS). The XPCS measures the decay of the density wave, $|\rho(\boldsymbol{q})|^2$, with time in the vicinity of the first peak of $S(Q)$, thus the decay of the MRO amplitude. This again suggests a link between the dynamical heterogeneity and the decay of the MRO.

### C. Atomic pinning of phason and amplitudon excitations

The density wave approach provides a natural description of atomic cooperativity in deformation from a point of view different from the existing theories. In the density wave picture a phason or amplitudon excitation affects all atoms within the volume of the window function. At high temperatures rapidly successive local atomic motions cause numerous phason and amplitudon excitations, reducing their effects on each atom to mere thermal noise. Thus, atoms do not interact each other through density wave exciations. In contrast, in deeply supercooled liquid well below $T_A$ and in glass below $T_g$, one phason or amplitudon excitation is well-separated in time from the next excitation, providing enough time for local cooperative atomic dynamics to occur for each excitation. To illustrate this point let us consider an isolated atom. In order to describe this atom by a density function, $\rho(r)$, we need to involve waves of all $q$s to form a $\delta$-function. However, when only the waves with a single value of $q = |q_{IG}|$ are used as in the current model, the $\rho_{IG}(r)$ centered at an atom is not a $\delta$-function, but it has many spherical layers of oscillating density just like the PDF, eq. (29), in addition to the peak at the center atom, as shown in Fig. 9. When the center atom is displaced, these spherical layers shift together and move other atoms nearby, resulting in cooperative motion within the MRO coherence volume. Thus, atomic cooperativity naturally emerges in the density wave picture.

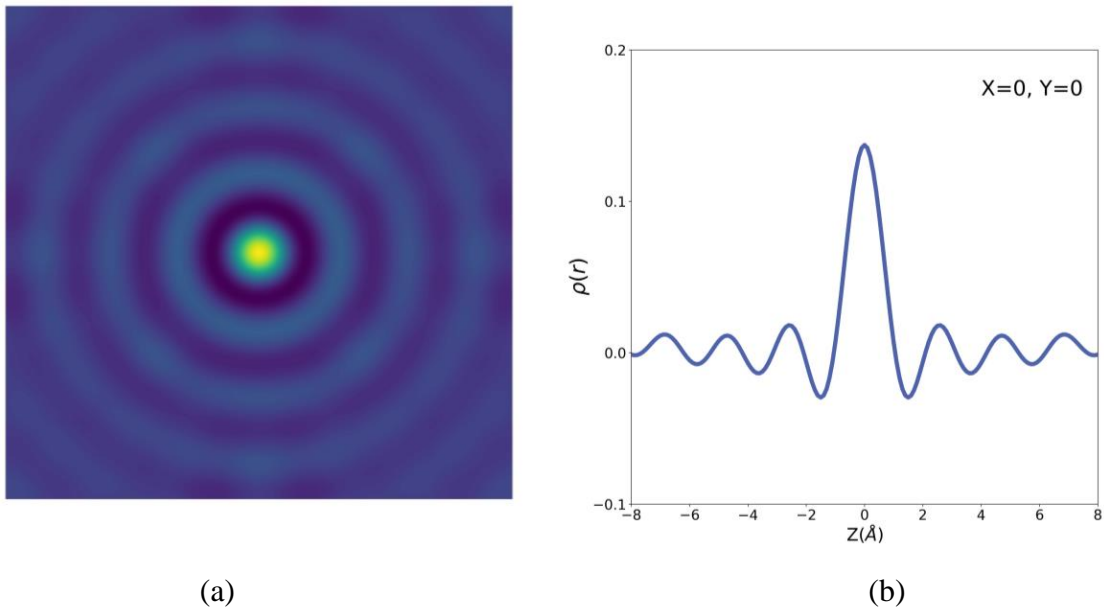

(a)           (b)

**Figure 9.** The $\rho(r)$ centered at a single atom at $r = 0$ obtained by overlapping density waves with $\delta_{IG} = 0$. (a) 2D color map at $z = 0$, and (b) a cut at $x = y = 0$.

There are two kinds of atomic displacement. One is thermal vibration by phonons, both propagating and localized, and the other is displacement which causes change in the structure by cutting or forming atomic bonds. Here atomic bond does not necessarily mean chemical bond. In the dense-random-packed (DRP) structure [40], atomic bond is defined between atoms which are nearest neighbors to each other. The second group of displacements change the local topology of atomic connectivity, so they are topological excitations [82 – 86]. In terms of the potential energy landscape (PEL) picture [4, 51, 87] a glassy system is trapped in a local minimum of the PEL, and structural change requires overcoming the PEL barrier. The second group of displacements cause changes in the inherent structure [4, 88]. For this reason, for the density waves to produce a move of the second kind and create local flow they have to move far enough to jump over the energy barrier. Unless such a jump is made the system is confined to a small space around a local PEL minimum. In other words, usually the amplitudes and phase factors are atomically pinned, unless phason and amplitudon excitations are large enough to jump over the PEL barrier. The activation energy of such a process must be proportional to the coherence volume, because atoms in the coherence volume move together cooperatively as discussed above. Interestingly, such a relation was observed for viscosity in $Pd_{42.5}Ni_{7.5}Cu_{30}P_{20}$ liquid just above $T_g$ [23], and for fragility which reflects viscosity [25].

We now discuss how the PEL traps density waves through atomic pinning of phasons and amplitudons. If we focus just on one atom and its neighbors, their PEL, local PEL, has minima at integer values of the coordination number, separated by energy barriers [89]. Whereas the concept of the local PEL is valid only above $T_A$, where cooperativity is absent [54], it provides a useful picture of local atomic pinning. Let us consider an atom surrounded by $N_C$ nearest neighbor atoms. In a gedanken experiment we expand the size of this atom at the center. At one time the shell of the nearest neighbors will become unstable, and an extra atom has to be added to the nearest neighbor shell to stabilize it. This is because the ideal value of $N_C$ increases with the size of the central atom relative to that of the surrounding atoms [90]. If the central atom is expanded to the size corresponding to the change in $N_C$ by more than 0.5, energetically it is more advantageous to add an extra atom to the nearest neighbor shell. Thus, $N_C$ is increased by one and the system slips into the next valley in the local PEL. For a system of atoms interacting with a harmonic or nearly harmonic potential, this instability condition can be expressed in terms of the critical volume strain [91, 92],

$$\varepsilon_V^{crit} = 0.11 . \tag{38}$$

This concept was successfully used in explaining the minimum solute concentration to stabilize metallic glass [93] and the glass transition temperature [94]. The atomic-level local volume fluctuations are conveniently described by the atomic-level pressure fluctuation [92, 95], which increases with temperature following the equipartition theorem [96, 97]. If the local volume strain in absolute value is smaller than $\varepsilon_V^{crit}$ this atom is trapped in one local PEL minimum. In this way the PEL pins phasons and amplitudons.

The energy-scale of the barrier height is given by the energy to change local coordination by cutting or forming atomic bonds. The most commonly found pattern of such change is the local topological change called Pachner move [85], involving about 5 atoms [68, 98]. A single Pachner move induces atomic displacements within the coherent volume, because atoms are connected by density waves as shown in Fig. 9. Thus, each Pachner move is dressed by cooperative motions of atoms of which number is proportional to the coherence volume, $V_c = \xi_s^3$. This argument explains why the activation energy for viscosity and diffusivity is proportional to $V_c$. When temperature is increased local density fluctuations increase [15]. The activation energy for viscosity, $E_a$, becomes decoupled from $V_c$, and becomes a constant above $T_A$. Details of this gradual transition is unclear, but it may be phenomenologically modeled by $E_a \sim \xi_s^\chi$, where the fractal dimension $\chi$ is equal to 3 at $T_g$ and decreases to zero at $T_A$ [24].

### D. Random-first-order-transition (RFOT) theory

In the random-first-order-transition (RFOT) theory [32 – 35] it is assumed that a liquid is made of mosaic of nano-scale domains of local states separated by interfaces between domains.

Each local state corresponds to a replica of the exact solutions in the infinite dimensions [5, 31], such as the Sherrington-Kirkpatrick model [99]. The change in the state is made through nucleation of a new domain, which costs energy to create new domain boundaries. This nucleation barrier results in kinetic delay just as in any first-order transition, which explains the relaxation of liquid structure. The picture of nano-scale domains of local states appears to be related to that of a liquid with overlapping windows of density wave states centered on each atom presented here. If we assume, although this is not proven, that the structurally coherent ideal glass density wave states discussed here correspond to the local states of the RFOT theory, non-overlapping domains separated by domain walls in the RFOT theory are produced by non-overlapping square window functions in eq. (34). Then the shape of the first peak of $S(Q)$ is proportional to $[W_{\xi,\Lambda}(\boldsymbol{q})]^2$, thus $\sim (q - q_{IG})^{-4}$, known as the Porod's law [100]. However, the observed Lorentzian peak shape of $S(Q)$ [23] is more consistent with the continuous overlapping states with exponential decay in space without domain walls, akin to the case of the second-order transition. In our model viscosity is produced by atomic pinning of the density waves, rather than the nucleation of a new state as in the RFOT theory.

### E. Hard-sphere model and jamming

The weak dependence of $q_{\min}$ on $\phi(r_c)$ as shown in Fig. 4 and the fact that similar results were obtained for three very different potentials suggest that the occurrence of a minimum in $\phi_{pp}(q)$ may be a consequence of geometry, rather than actual details of the shape of $\phi(r)$. In fact, even the hard-sphere (HS) model has a minimum in the pseudopotential, if we introduce an energy cutoff rather than an infinite repulsive potential [18]. It is possible that the mere action of eliminating atomic overlaps by the HS potential results in a preferred set of density waves through the resonant interference between the scattered and incoming density waves. The similarity between the HS diameter and the wavelength of the density wave at the minimum of $\phi_{pp}(q)$ [18] strongly suggests such a scenario. Of course, the total energy still is positive (repulsive) because the energy minimum in $\phi_{pp}(q)$ is shallower than the potential energy for uniform density. Interestingly, the long-range oscillation was observed in the hard-sphere model [101] which may be related to this minimum. Even though the focus of Ref. 101 is on proving hyperuniformity of the jammed state, the appearance of long-range oscillations in the PDF is striking, demanding explanation. In Ref. 101 the jammed HS model was created by using a linear potential rather than a true hard potential to achieve full relaxation. This particular procedure of production may have incurred the pseudopotential effect. Thus, the instability of a uniform high-temperature gas state against the density wave sate could be prevalent once the interatomic potential is introduced, irrespective of details of the potential. In the HS model resistance for deformation is provided by jamming [30, 31]. Here jamming is expressed by the stiffness of the density waves which creates an energy barrier.

### V. Conclusion

The conventional approaches to elucidate the atomic structure of liquid and glass are *bottom-up* approaches, in which one starts with focusing an atom and its neighboring atoms and adds more atoms to form the global structure, locally minimizing the interatomic potential energy. However, with this approach it is difficult to comprehend why well-defined MRO oscillations are always observed, even in metallic liquids with complex compositions. It is commonly assumed that the MRO is a direct consequence of the SRO. But the SRO and MRO are fundamentally different in nature, because the SRO describes atomic correlations within the atomic cage, whereas the MRO describes correlations with coarse-grained density fluctuations. To resolve this conundrum, we suggest that the MRO is not a direct consequence of the SRO but is driven by a different collective force. We propose a two-way approach composed of an inverted *top-down* approach as well as the bottom-up approach to conceptualize the formation of the structure. In the top-down approach we start with a high-density gas state and minimize the global potential energy through density waves which minimizes the pseudopotential energy. The two approaches are not fully compatible, and the competition and compromise between the two driving forces result in a structure with the MRO. The strength of the MRO, which is expressed by the MRO coherence length and is called "ideality", is unrelated to the atomic size mismatch and compositional complexity [52]. It is suspected that the strong driving force for the SRO, such as strong covalency, compromises the MRO. The MRO coherence length is directly related to liquid fragility [25] and viscosity near $T_g$ [23]. This is because in supercooled liquid the density waves are pinned to atoms through the phase factor and amplitude (phason and amplitudon), and control atomic dynamics through the coherence volume. The even-handed two-way approach proposed here provides a more balanced account of the complex structure and dynamics in liquids and glasses. This could form a basis for developing a more realistic and general theory to describe the state of liquid and glass.

**Acknowledgment**

The authors thank M. L. Manning, A. Lemaitre, Y. Shinohara, C. K. C. Lieou and E. Zarkadoula for helpful discussions. This work was supported by the U. S. Department of Energy, Office of Science, Basic Energy Sciences, Materials Sciences and Engineering Division. This research used resources of the National Energy Research Scientific Computing Center (NERSC), a U.S. Department of Energy Office of Science User Facility located at Lawrence Berkeley National Laboratory, operated under Contract No. DE-AC02-05CH11231 using NERSC award BES-ERCAP0020503. This manuscript has been authored by UT-Battelle, LLC under Contract No. DE-AC05-00OR22725 with the U.S. Department of Energy. The United States Government retains and the publisher, by accepting the article for publication, acknowledges that the United States Government retains a non-exclusive, paid-up, irrevocable, world-wide license to publish or reproduce the published form of this manuscript, or allow others to do so, for United States Government purposes. The Department of Energy will provide public access to these results of federally sponsored research in accordance with the DOE Public Access Plan (http://energy.gov/downloads/doe-public-access-plan).

**Appendix A**

A measure of the magnitude of $k_B T_u = \phi(r_c)$ may be given by the timescale of overcoming the potential,

$$\tau_{pp} = \tau_D \exp\left(\frac{\phi(r_c)}{k_B T}\right), \tag{A1}$$

where $\tau_D$ is the Debye time (~ $10^{-13}$ sec). This should be larger than the total MD run time, $\tau_{MD}$. For liquid Fe at 2000K for $\tau_{MD} = 10^{-9}$ sec, $\phi(r_c)$ must be larger than 2 eV.

**Appendix B**

The Lennard-Jones (LJ) potential,

$$\phi_{LJ}(r) = A\left[\left(\frac{r}{\sigma}\right)^{12} - \left(\frac{r}{\sigma}\right)^{6}\right], \tag{A2}$$

with $A = 0.1$ eV and $\sigma = 2.33$ Å, and the Yukawa potential,

$$\phi_Y(r) = \frac{B}{r}\exp(-\kappa r), \tag{A3}$$

with $B = 10^5$ eV and $\kappa = 5.328$ Å$^{-1}$, were used in simulation.

**Appendix C**

From eqs. (11) and (12), for an atomic system,

$$\rho(\mathbf{q}) = \frac{1}{V}\sum_i e^{-i\mathbf{q}\cdot\mathbf{R}_i}. \tag{A4}$$

Thus, the structure function $S(\mathbf{q})$ is given by

$$S(\mathbf{q}) = \frac{1}{N}\sum_{i,j} e^{-i\mathbf{q}\cdot(\mathbf{R}_i - \mathbf{R}_j)} = \frac{V}{\rho_0}|\rho(\mathbf{q})|^2. \tag{A5}$$

For a spherical system,

$$S(q) = \langle S(\mathbf{q})\rangle = \frac{1}{4\pi}\int S(\mathbf{q})d\mathbf{\Omega}_q = \frac{V}{\rho_0}\langle|\rho(\mathbf{q})|^2\rangle \tag{A6}$$

For a Bragg peak at $q = q_1$, the area of the peak is,

$$A_\rho(q_1) = \frac{V}{\rho_0}\langle|\rho(\mathbf{q}_1)|^2\rangle \Delta q, \tag{A7}$$

where

$$\Delta q = \frac{2\pi}{L}. \tag{A8}$$

Then,

$$A_\rho(q_1) = \frac{2\pi}{\rho_0} B_\rho(q_1), \tag{A9}$$

Thus,

$$\begin{aligned} g(r) - 1 &= \frac{1}{2\pi^2 \rho_0 r} \int [S(Q) - 1] Q \sin(Qr) dQ \\ &= \frac{A_\rho(q_1)}{2\pi^2 \rho_0 r} q_1 \sin(q_1 r) = B_\rho(q_1) \frac{q_1^2}{\pi \rho_0^2} \frac{\sin(q_1 r)}{q_1 r} \end{aligned} \tag{A10}$$